# Probing supermassive black hole mergers and stalling with pulsar timing arrays

Chiara M. F. Mingarelli

**The observation of gravitational-waves from merging supermassive black holes will be transformative: the detection of a low-frequency gravitational-wave background can tell us if and how supermassive black holes merge, inform our knowledge of galaxy merger rates and supermassive black hole masses, and enable the possibility of detecting new physics at nanohertz frequencies. All we have to do is time pulsars.**

Supermassive black hole (SMBH) mergers are the strongest sources of gravitational-waves (GWs) in the Universe. SMBH mergers are expected to follow galaxy mergers, since likely all massive galaxies host central SMBH, e.g. [1]. Briefly, the black holes fall to the center of the newly-formed galaxy through dynamical friction and form a binary. This binary should harden by ejecting stars crossing their orbit – a process called stellar hardening. When the binary is separated by a centiparsec to a milliparsec, gravitational waves (GWs) drive the binary to merge. These steps are described in more detail in [2].

However, there is currently sparse observational evidence for sub-parsec separated SMBHB systems. In fact, under the assumption of a static spherical galaxy model, SMBHBs may stall at their final parsec of separation, and never merge – this is known as the final parsec problem [3]. This stalling happens when, for example, the black holes run out of stars to eject – called loss cone depletion. Even though the system is emitting GWs, it would take longer than a Hubble time for the binary to merge via GW emission only. However, if one assumes a triaxial galactic model, loss cone depletion is no longer an issue [4].

There are other ways to overcome the final parsec problem: a circumbinary accretion disk can torque the binary, carrying away energy and angular momentum, helping it to overcome its final parsec of separation and merge. Environmental interactions with gas and stars also induce eccentricity in the binary [5]. An eccentric binary will emit GWs at higher harmonics, rapidly decreasing the time to coalescence of the binary, and enabling it to merge within a Hubble time.

## Pulsar Timing Arrays
A pulsar timing array (PTA) is a galactic-scale GW detector. With an array of millisecond pulsars, one can search for nanohertz frequency GW signals originating from the most massive SMBHB systems in their slow inspiral phase, Refs [6,7]. GWs transiting the galaxy change the proper distance between the pulsars and the Earth by tens of meters per light year, inducing delays or advances in the pulsar's pulse arrival times. These time delays are tens to hundreds of nanoseconds over a decade, highlighting the importance of millisecond pulsars as the basis of our experiment: no other natural object has this kind of timing precision.

The nanohertz GWs in the PTA band likely originate from SMBHBs in the $10^8 - 10^9\ M_\odot$ range, with periods of years to decades, hence, they are millions of years from merging. A nanohertz GW background should be generated from the cosmic merger history of SMBHs and will likely be detected by PTAs in the next three to five years [8]. Assuming circular binaries which have decoupled from their environment (and there therefore merging due to GW emission), one can write the characteristic strain of the background as a function of its amplitude at a reference frequency of 1/yr:

$$h_c = A \left(\frac{f}{yr^{-1}}\right)^\alpha$$ Equation 1,

where $\alpha = -2/3$ for black holes [9], Figure 1. The amplitude of the GW background depends on the black hole masses, the merger timescales and the pair fraction. While we have written the characteristic strain as a power-law in Equation 1, this only holds true if the binaries are circular. In fact, the shape of the strain spectrum depends on how SMBHs merge [5], Figure 1, which we explore below.

**The GW Background**
The solution to the final parsec problem can leave an imprint in the strain spectrum (Equation 1) of the GW background. Since stellar hardening, torques from a circumbinary disk, and eccentricity are effective ways of shrinking the binary at parsec-scale separations – much more effective than gravitational radiation -- they can deplete the power in the GW background's very low frequencies (wide binary separations), Figure 1. The transition from environmental interactions dominating the binary evolution to GW emission happens at a few nanohertz: roughly 3nHz for stellar hardening, and 0.14 nHz for accretion models, e.g. [10] and references therein. One can therefore search for a turnover in the strain spectrum of the background and constrain e.g. the density of stars surrounding the binary, causing it to merge via stellar hardening. This was done for the first time in ref. [10].

In the unlikely scenario where some fraction or all binaries stall, interactions between multiple SMBHs from subsequent galaxy mergers with mass ratios $10^{-4} < q_* < 1$ will allow the black holes to merge. This creates a floor to the GW background, albeit at a factor of a few lower than the amplitude of models with no stalling [11,12].

**The next 10 years**
SMBHBs merge before ever reaching the high-frequency LIGO band. For example, the innermost stable circular orbital frequency of an equal-mass $10^9\ M_\odot$ SMBHB is ~4$\mu$Hz, assuming it is not spinning. This is also outside the range of GW frequencies accessible by the future LISA mission, scheduled to launch in 2034. The LISA instrument is instead expected to detect $10^5 - 10^6\ M_\odot$ SMBHBs, as well as inspiralling extreme mass ratio BH systems and a foreground from galactic white dwarf binaries.

A $3\sigma$ detection of the nanohertz GW background is currently possible with PTAs, and likely to occur in three to five years, with the details depending on the underlying astrophysics of SMBH mergers, Ref [8]. Figure 2 shows the time to detection of the GWB given the "true" amplitude of the background, at a reference frequency of 1/yr, assuming a power-law model. This study included the current IPTA pulsars' noise properties and added six pulsars per year with 250 ns timing residuals, labelled IPTA+.

We are entering into an exciting detection and non-detection era alike: current upper limits disfavor the most optimistic models of the GW background [13] and are starting to be in tension with moderate models, e.g. Ref [5]. Indeed, since the GW background signal is expected to build up slowly over time [14], one expects to see a 2σ level feature at least one or two years before a $3\sigma$ detection (X. Siemens, private communication). If detection prospects increase as expected, in ~5 years one could detect a GW background with an amplitude of $A = 5 \times 10^{-16}$, enabling us to probe a GW background generated from stalled SMBHBs, merging through subsequent many-body BH interactions [11,12]. In 6 years, one may be able to determine if in fact supermassive black hole masses have really been overestimated [15]. Importantly, both [11] and [12] show that there is a floor to the GWB at around $A = 10^{-16}$. According to the results from [8], it should take PTAs another 10 years to reach this floor, Figure 2.

Five years after the detection of the nanohertz GW background, individual SMBHB systems are expected to be detected [16]. These individual binary systems will be fascinating to study: if an electromagnetic counterpart is detected, multimessenger astrophysics will be possible until the binary merges, which can be anywhere from 150 yrs to 25 Myrs. Ten years after the GW background detection, we expect to characterize it sufficiently well to extract its underlying anisotropy [14, 18], which should follow large-scale structure.

Importantly, PTA experiments should hit the floor of the GW background years before LISA flies in 2034. Merging $10^5 - 10^6 \, M_\odot$ SMBHs in the LISA band are likely the progenitors of the $10^8 - 10^9 \, M_\odot$ SMBHs inspiralling in the PTA band, and studies are underway to understand what a PTA non-detection of the GW background and individual SMBHB sources means for LISA. Since even stalled SMBHs can eventually merge through many-body interactions, this may point to a lower-than expected SMBH population fraction.

Indeed, the next decade of PTA data has the potential to transform what is known about massive SMBHs mergers, the SMBH population fraction, the relationship between SMBH mass and galaxy mass, galaxy merger rates, and fundamental physics at nanohertz frequencies. Both the detection and non-detection of the GW background in the next 5 years will signal the beginning of this exciting era: all we have to do is time pulsars.

---

*Chiara M. F. Mingarelli is at the Center of Computational Astrophysics at the Flatiron Institute, 162 Fifth Ave, New York, NY 10010, USA.*

**Acknowledgements**
CMFM thanks J. Lazio, D. Foreman-Mackey, X. Siemens for useful discussions. CMFM thanks S. Taylor and S. Burke-Spolaor for permission to edit and reproduce some of their figures. The Flatiron Institute is funded by the Simons Foundation.

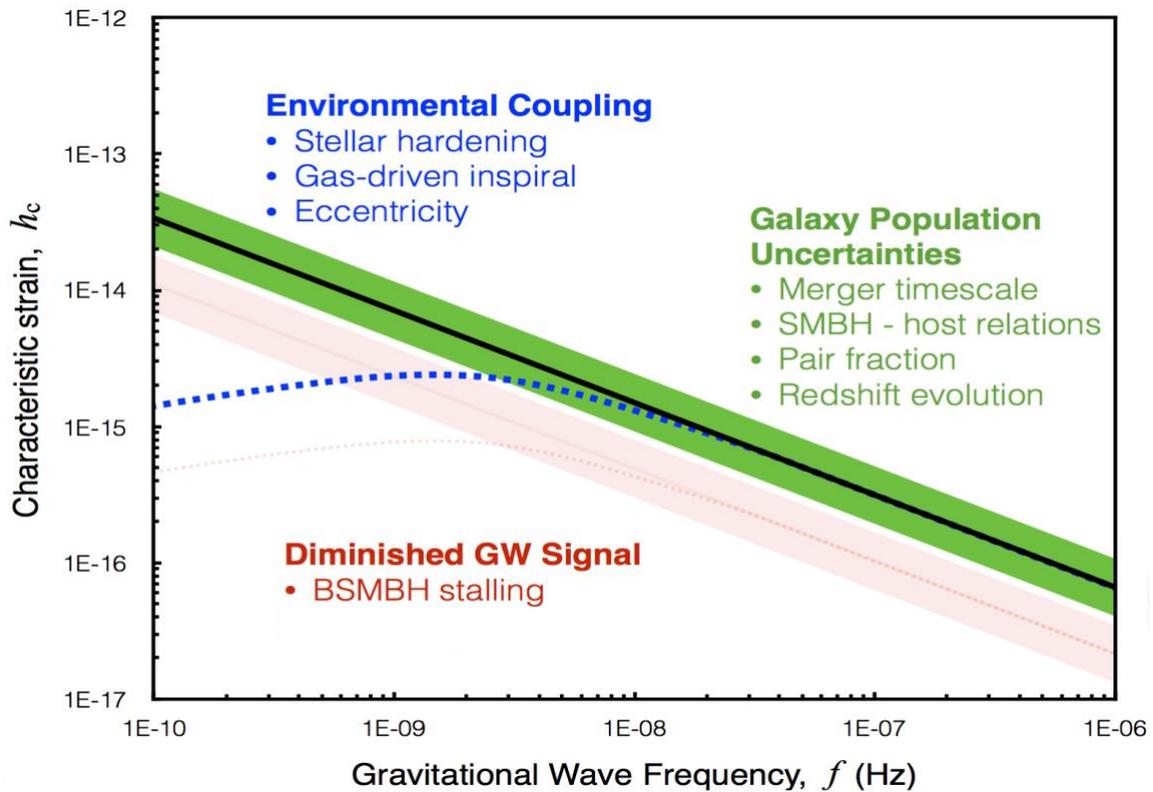

**Figure 1: Astrophysics manifesting in the gravitational-wave background strain spectrum.** The black line is the gravitational-wave background (GWB) strain spectrum assuming circular SMBHBs which have decoupled completely from their environment. The green shaded area represents galaxy population uncertainties which affect the amplitude of the strain spectrum: SMBHB merger timescales, uncertainties in SMBH masses (drawn from supermassive black hole – host galaxy scaling relations), the SMBH pair fraction, and the redshift evolution of these uncertainties. The blue dotted line is the resulting shape of the strain spectrum if the final parsec problem is over-solved: either via stellar hardening, circumbinary disc interactions, and eccentricity. These are all more effective at carrying away energy from the binary at wide separations (or low frequencies) than GW emission. The black holes could also stall, pink region, which would lead to an overall decrease in the amplitude of the strain spectrum. Figure credit: S. Burke Spolaor, Ref [19].

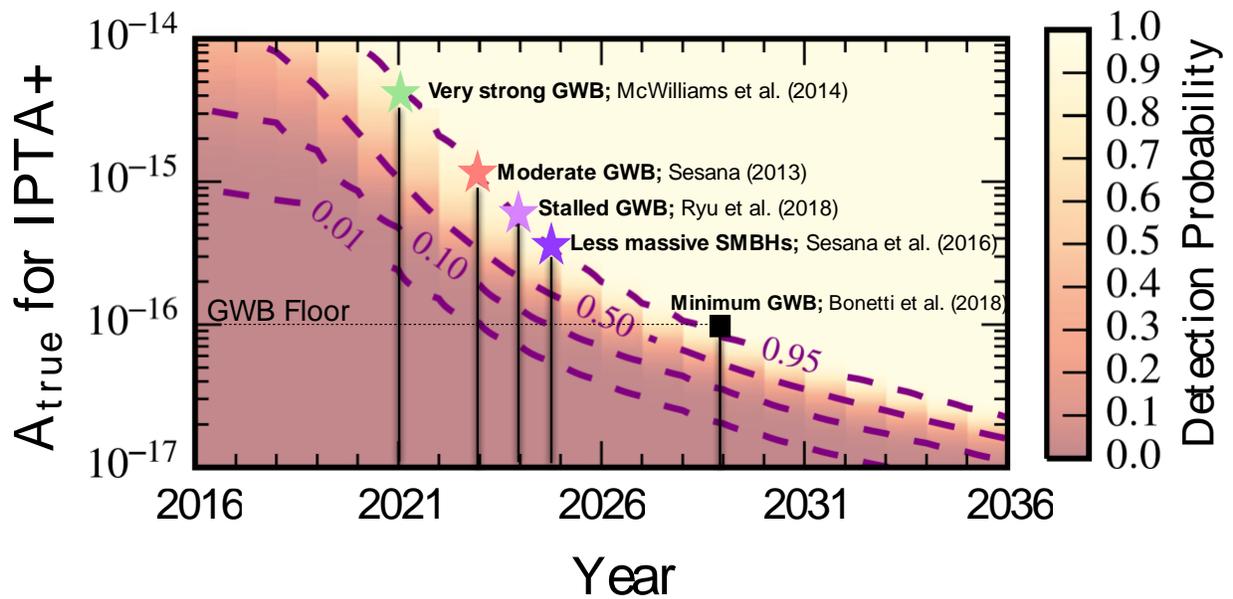

**Figure 2:** Time to detection of different models of the gravitational-wave background (GWB). The dashed lines are the detection probability contours, and the labels show the mean value of the GWB amplitude at a reference frequency of 1/yr. For example, in 2021 there is a 95% chance of making a $3\sigma$ detection of the GWB predicted by Ref [13]. Ref [13] is the most optimistic model, and is in tension with current upper limits, while Ref [14] is starting to be in tension with upper limits. The amplitude of a GWB arising from a population of stalled binaries, eventually merging through many-body interactions, will be detectable in 5 years (Ref [11]), and if supermassive black hole masses are biased large (Ref [15]), the background will be detected in 6 years. In the worst-case scenario, explored by Ref [12], we will hit the floor of the GWB in 10 years from now. Figure modified with permission from S. Taylor from Taylor et al. (2016) [8].